\documentclass[onecolumn,10pt]{article} 

\usepackage[square,numbers,sort&compress,comma]{natbib}
\usepackage{amsmath}
\usepackage{amssymb}
\usepackage{caption}
\usepackage{graphicx}
\usepackage{latexsym}
\usepackage{times}
\usepackage[pagewise]{lineno}
\usepackage{xcolor}

\topmargin - 12pt 
\renewenvironment{abstract}%
              {
               \small
               {\bfseries \abstractname}
               \par
               \vspace{10pt}
             }

\renewcommand\abstractname{Abstract}

\newcommand{\nomenclature}
              [1]
              {
               \bgroup
               \flushleft
               \small\bf
               #1
               \par
               \egroup
             }

\renewcommand{\section}
              [1]
              {
               \bgroup
               \flushleft
               \small\bf
               \refstepcounter{section}
               \arabic{section}. #1
               \par
               \egroup
             }

\renewcommand{\subsection}
              [1]
              {
               \bgroup
               \flushleft
               \small\em
               \refstepcounter{subsection}
               \arabic{section}.
               \arabic{subsection}. #1
               \par
               \egroup
             }

\renewcommand{\subsubsection}
              [1]
              {
               \bgroup
               \flushleft
               \small\em
               \refstepcounter{subsubsection}
               \arabic{section}.
               \arabic{subsection}.
               \arabic{subsubsection}. #1
               \par
               \egroup
             }

  \newcommand{\acknowledgement}
              [1]
              {
               \bgroup
               \flushleft
               \small\bf
               #1
               \par
               \egroup
             }

  \newcommand{\sectionbib}
              [1]
              {
               \bgroup
               \flushleft
               \small\bf
               #1
               \par
               \egroup
             }

\begin{document}

\title{\LARGE Real-fluid Transport Property Computations Based on the Boltzmann-weighted Full-dimensional Potential Model}

\author{{\large Xin Zhang$^{a}$, Junfeng Bai$^{c}$, Bowen Liu$^{a}$, Tong Zhu$^{b, *}$, Hao Zhao$^{a, *}$}\\[10pt]
        {\footnotesize \em $^a$College of Engineering, Peking University, Beijing, 100871, China}\\[-5pt]
        {\footnotesize \em $^b$School of Chemistry and Molecular Engineering, East China Normal University, Shanghai, 200062, China}\\[-5pt]
        {\footnotesize \em $^c$Department of Mechanical Engineering, City University of Hong Kong, 999077, Hong Kong}\\[10pt]
        {\footnotesize \em $^*$Corresponding author: Hao Zhao (h.zhao@pku.edu.cn), Tong Zhu (tzhu@lps.ecnu.edu.cn)}}
\date{}


\small
\baselineskip 10pt


\vspace{50pt}
\maketitle
\vspace{40pt}
\rule{\textwidth}{0.5pt}
\begin{abstract} 

The intermolecular potential plays crucial roles in real-fluid interactions away from the ideal-gas equilibrium, such as supercritical fluid, high-enthalpy fluid, plasma interactions, etc. We propose a Boltzmann-weighted Full-dimensional (BWF) potential model for real-fluid computations. It includes diverse intermolecular interactions so as to determine the potential well, molecular diameter, dipole moment, polarizability of species without introducing bath gases, allowing more accurate descriptions of potential surfaces with more potential parameters. The anisotropy and temperature dependence of potential parameters are also considered by applying the Boltzmann weighting on all orientations. Through the high-level Symmetry-Adapted Perturbation Theory calculations, full-dimensional potential energy surface datasets are obtained in 432 orientations for each species. Subsequently, the Boltzmann-weighted Full-dimensional potential parameters are derived by training the dataset exceeding $5*10^6$ data, including nonpolar and polar molecules, radicals, long-chain molecules, and ions. These BWF transport properties calculated by the BWF potential have been compared against the Lennard-Jones transport properties as well as experimental viscosity, mass diffusivity, and thermal conductivity coefficients. It shows discrepancies of viscosity coefficients within 1\% and 5\% for nonpolar and polar molecules, respectively. Furthermore, this potential model is applied to study radicals, long-chain molecules, and ions, for which the experimental data is rarely accessed in high accuracy. It indicates significant prediction improvements of complex interactions between various particles. The new transport properties are also embedded to predict the laminar flame speeds and the flame extinction limits of methane, dimethyl ether, and n-heptane at elevated pressures, confirming its predictivity and effectiveness. 

\end{abstract}
\vspace{10pt}
\parbox{1.0\textwidth}{\footnotesize {\em Keywords:} Real-fluid; Transport property; Intermolecular potential; Non-equilibrium simulation; Supercritical combustion}
\rule{\textwidth}{0.5pt}
\vspace{10pt}

\clearpage


\vspace{20pt}

{\bf 1) Novelty and Significance Statement}
\vspace{10pt}

Intermolecular interactions are significantly amplified in nonideal flows. These flows are highly relevant to propulsion and energy conversion processes. Unfortunately, the real-fluid intermolecular interactions in the literature are mainly based on the Lennard-Jones potential, which reveals diverging errors for transport predictions and reactive flow simulations at high to ultra-high pressures, especially for long-chain molecules, radicals, and ions. We present a Boltzmann-weighted Full-dimensional potential model, considering various intermolecular interactions and including the anisotropy and temperature dependence of potential parameters. It is applied to compute transport properties of long-chain molecules, radicals, and ions, and also applied to combustion simulations of different fuels with high accuracies, providing comprehensive and robust supports in transport libraries and reactive flow simulations.

\vspace{20pt} 

{\bf 2) Author Contributions}
\vspace{10pt}

\begin{itemize}

        \item{Xin Zhang designed research, performed research, analyzed data, wrote the paper;}

        \item{Bowen Liu designed research, analyzed data;}

        \item{Junfeng Bai designed research, analyzed data;}
        
        \item{Tong Zhu designed research;} 

        \item{Hao Zhao designed research.} 
\end{itemize}

\vspace{20pt}



\clearpage


\section{Introduction} \addvspace{10pt}

Real fluids, such as supercritical fluid, high-enthalpy fluid, and plasma, play important roles in propulsion and energy conversion scenarios: (i) supercritical spray, active cooling, and combustion in advanced combustion engines and rockets, (ii) thermal protection of hypersonic aircrafts and reentry of satellites, (iii) plasma-assisted propulsion, ignition, synthesis, and material treatment, etc. In such processes, intermolecular interactions are dramatically amplified. For example, the supercritical combustion takes place when the fuel and the oxidizer interact at the supercritical phase at elevated pressures and temperatures. Due to the shorter molecular distance and faster collisions in the intense interactions of supercritical fluids, it provides a higher thermodynamic efficiency and lower emissions in advanced internal combustion engines and rockets \cite{super1, super2,super3,super4}. Another example is plasma-assisted combustion and synthesis, where the highly non-equilibrium reactive flow is created in plasma with increased ionizations and excitations of molecules, radicals, and ions. These energetic species accelerate ignitions, combustions, and material synthesis in a non-equilibrium way under an intense intermolecular force field \cite{plasma1, plasma2, plasma3}. 

The real-fluid effects come from strong intermolecular interactions, manifesting in the compressibility, thermodynamic, and transport properties, which further affect combustion characteristics, such as the laminar flame speed and the flame extinction limit. Bai \cite{bai} computed the compressibility and thermodynamic properties of nonpolar and polar molecules by employing the Virial equation of state (EoS) \cite{text} based on the Lennard-Jones potential \cite{LJ} and the Stockmayer potential \cite{Stockmayer}. It is found that the predictions incorporating real-fluid effects significantly outperform the empirical Redlich-Kwong (RK) EoS \cite{RK} below 200 atm. 

At the same time, accurate transport properties are essential in the combustion modeling \cite{brown2,Wang4, Esposito1}. Brown \cite{brown-re} and Jasper \cite{Jasper-re} extensively reviewed the methods for estimating transport properties. For instance, Stallcop \cite{stallcop1, stallcop2, stallcop3, stallcop4} and Wang \cite{Wang1, Wang2, Wang3} provided the collision cross sections and diffusion coefficients of binary collision systems, such as H/N\textsubscript{2} and H/Ar, using the first-principle method. Dagdigian \cite{dag1,dag2,dag3} has performed detailed calculations of the transport properties of H/CO, H/CO\textsubscript{2}, and H\textsubscript{2}O/H collision pairs using the quantum scattering method. Violi \cite{Violi1, Violi2} used the molecular dynamics simulation to determine the diffusion coefficients of long-chain alkanes and revealed the effect of the molecular configuration on the diffusion coefficients. Patidar \cite{Patidar} used the advanced Ab-initio method to provide potential energy surfaces and transport coefficients of a variety of high-energy organic molecules with four bath gases. 

Presently, most of the calculations for transport properties above are based on the Lennard-Jones potential \cite{LJ} and the Stockmayer potential \cite{Stockmayer} (hereafter both potentials are referred to as the LJ potential), which are used by CHEMKIN \cite{CHEMKIN} and Cantera \cite{Cantera} simulations and have been the dominant force fields in the field of combustion. However, the LJ potential has some shortcomings. Jasper \cite{Jasper1,Jasper2} studied the application of the LJ potential in the transport property calculations in details. It was found that significant transport errors were introduced due to the isotropic approximation, and the description of repulsive interaction. (i) The anisotropy, though negligible in small-scale systems such as H/N\textsubscript{2} and H\textsubscript{2}/N\textsubscript{2}, can exhibit a significant effect for the larger species such as n-C\textsubscript{4}H\textsubscript{10}/N\textsubscript{2} at lower temperatures, resulting in a 15\% error in the diffusion coefficient computations. Violi \cite{Violi1,Violi2} also emphasized the significance of the anisotropy on the transport properties, particularly for large molecules. (ii) Moreover, the incomplete description of the repulsive wall is a significant source of errors, introducing up to 40\% errors by using a steeper repulsive wall in the LJ potential than Buckingham (exp/6) potential \cite{buch}. Therefore, the LJ potential has good predictions for the simple nonpolar molecule systems, but is often unsatisfactory for long-chain molecules. In addition, the inert gas is required to be introduced as the bath gas when calculating the transport properties of polar molecules utilizing the LJ potential, whereas the mixing rules introduce extra errors \cite{brown1}. For the above considerations, there is an urgent need to develop a new potential model with accurate and comprehensive descriptions of intermolecular interactions at high pressures.

In this study, we will develop a Boltzmann-weighted Full-dimensional (BWF) potential model for the real fluids. This model takes into account various intermolecular interactions including the repulsive, induction, dispersion, and electrostatic interactions in details. The anisotropy and temperature dependence of potential parameters are also considered by applying the Boltzmann weighting on all intermolecular orientations. Subsequently, we train the Boltzmann-weighted Full-dimensional potential parameters across nonpolar and polar molecules, radicals, long-chain molecules, and ions by the dataset from the high-level Symmetry-Adapted Perturbation Theory \cite{SAPT} calculations. Then, the Boltzmann-weighted Full-dimensional transport properties calculated by the BWF potential parameters are compared with the LJ transport properties in literatures and the National Institute of Standards and Technology (NIST) database \cite{NIST}. Finally, we have embedded the new transport properties into combustion simulations, and calculated laminar flame speeds and flame extinction limits for various fuels, such as methane, dimethyl ether, and n-heptane.

\section{Materials and methods} \addvspace{10pt}

\subsection{Transport property calculation} \addvspace{10pt}

The viscosity coefficient $\eta$, mass diffusivity coefficient $D$, and thermal conductivity coefficient $\lambda$ are developed from the kinetic theory using classical mechanics of binary collisions as described by Hirschfelder \cite{text} (Eqs. 1-3). Potential parameters are used to calculate the transport properties of each species at different temperatures and pressures.

\begin{equation}
        \eta=\frac{5}{16} \frac{\sqrt{\pi m k_B T}}{\pi \sigma^2 \Omega^{(2,2) *}} 
        \end{equation}
\begin{equation}
        D=\frac{3}{16} \frac{\sqrt{\frac{2 \pi k_B{}^3 T^3}{m}}}{p \pi \sigma^2 \Omega^{(1,1) *}} \\
        \end{equation}
\begin{equation}
        \lambda=\frac{\eta}{w}\left(f_{\text {trans}} C_{v, \text {trans}}+f_{\text {rot}} C_{v, \text {rot}}+f_{\text {vib}} C_{v, v i b}\right)
        \end{equation}
where $ k_B, m, T, w, \sigma, \ p, and \ \Omega^{(l,s) *}$ are the Boltzmann constant, mass of the molecule, temperature, molecular weight, molecule diameter, pressure, and the collision integral, respectively. 
Here the individual species conductivities are composed of translational, rotational, and vibrational contributions (Eq. 4).

\begin{equation}
\begin{aligned}
        &\begin{gathered}
        f_{\text {trans}}=\frac{5}{2}\left(1-\frac{2}{\pi} \frac{C_{v, \text {rot}}}{C_{v, \text {trans}}} \frac{A}{B}\right), C_{v, \text {trans}}=1.5 R \\
        f_{\text {rot}}=\frac{\rho D}{\eta}\left(1+\frac{2}{\pi} \frac{A}{B}\right), C_{v, \text {rot,nonlinear}}=1.0 R, C_{v, \text {rot,linear}}=1.5 R
        \end{gathered}\\
        &\begin{gathered}
        f_{v i b}=\frac{\rho D}{\eta}, \quad C_{v, v i b}=C_v-C_{v, \text {trans}}-C_{v, \text {rot}} \\
        A=\frac{5}{2}-\frac{\rho D}{\eta}, B=Z_{\text {rot}}+\frac{2}{\pi}\left(\frac{5}{3} \frac{C_{v, \text {rot}}}{R}+\frac{\rho D}{\eta}\right)
        \end{gathered}
        \end{aligned}
        \end{equation}

However, Hirschfelder et al. \cite{text} pointed out that the transport properties calculated using the above formulas do not consider the pressure dependence. The transport properties are solely functions of temperature. 
Therefore, based on the Enskog theory \cite{text} (Eqs. 5-8), we computed the correction factors for the BWF model and the LJ model to obtain corrected transport coefficients over a wide range of temperatures and pressures.

\begin{equation}
        \frac{\eta \tilde{V}}{\eta^0 b_0}=\frac{1}{y}+0.8+0.761 y 
        \end{equation}
\begin{equation}
        \frac{\lambda \tilde{V}}{\lambda^0 b_0}=\frac{1}{y}+1.2+0.755 y 
        \end{equation}
\begin{equation}
        \frac{p D}{(p D)^0} \frac{\tilde{V}}{b_0}=\frac{1}{y}+1
        \end{equation}
where the superscript of 0 indicates transport properties at atmospheric pressure, $\tilde{V}$ is the molar volume of a substance, and the correction factor y is from the partial derivation of the equation of state. 
Here, we use the Virial coefficient equation of state and calculate the $2^{nd}$ and $3^{rd}$ virial coefficients from the BWF and the LJ potential models based on the statistical mechanics  \cite{text}. 
Then we obtain the Virial equation of states and the correction factors y for each species.

\begin{equation}
        y=\frac{\tilde{V}}{R T}\left[T\left(\frac{\partial p}{\partial T}\right)_{\tilde{V}}\right]-1
        \end{equation}
        
\subsection{Theory and model} \addvspace{10pt}

In the field of the molecular dynamics, a profound comprehension of the intermolecular interaction is crucial for predicting the behavior of combustion systems. For simple nonpolar gases, which only involve the repulsive and dispersion interactions, simple potential models, such as the LJ potential, can give a good description. However, in more complex molecular systems, various types of interactions come into play. 

\begin{figure*}[ht!]
        \centering
        \vspace{0.0 in}
        \includegraphics[width=300pt]{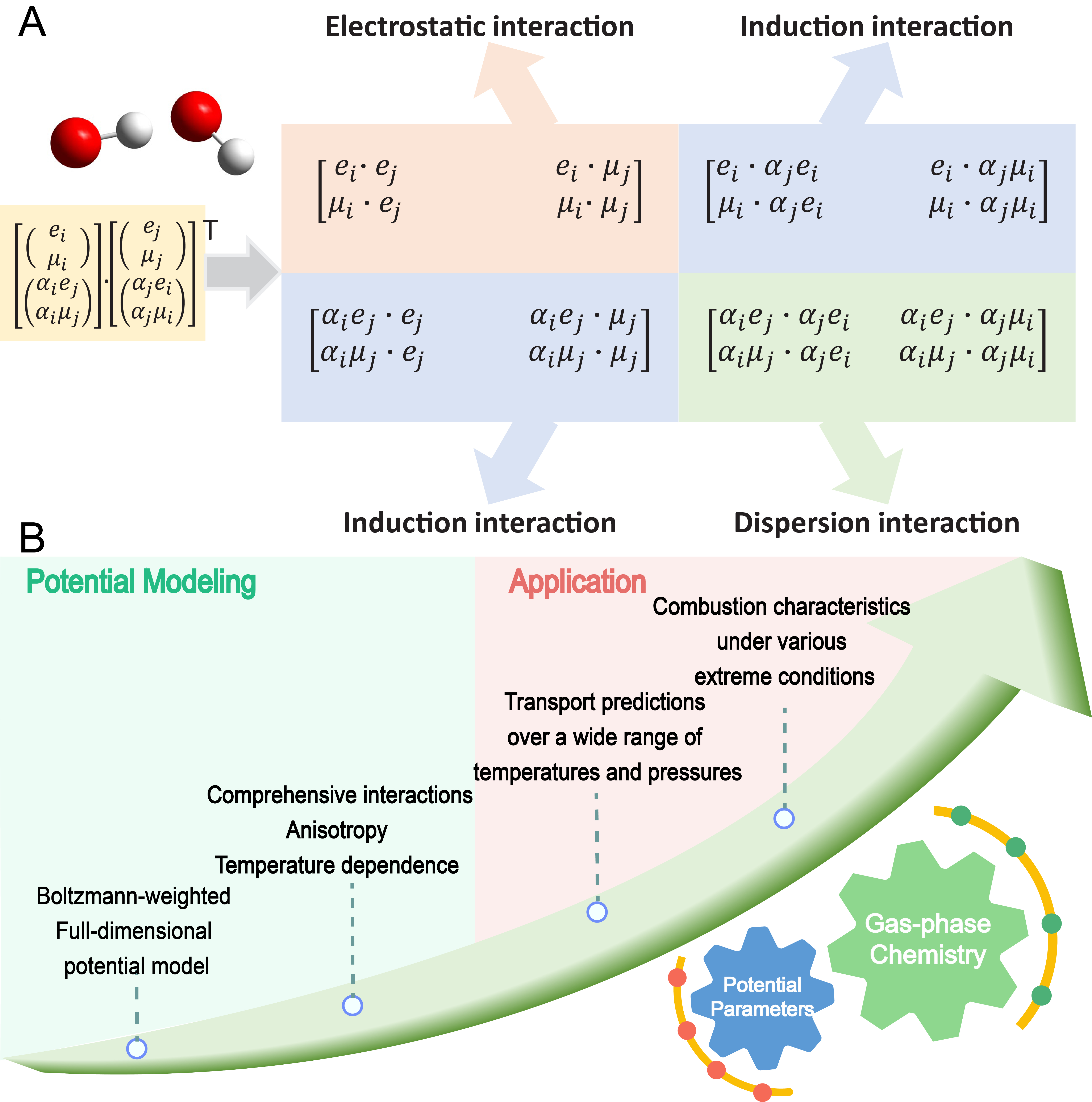}
        \vspace{5 pt}
        \caption{(A) The second-order tensor interaction space between molecules; (B) The diagram of real-fluid effects in the reactive flow simulations based on the Boltzmann-weighted Full-dimensional potential model.}
        \label{flow}
        \end{figure*}

To delve into the diverse interactions between molecules, we scrutinize their behavior based on the distance between the centers of particles. When two molecules are in close proximity, their electron clouds will overlap. According to the Pauli exclusion principle \cite{pauli}, two identical fermions cannot occupy the same quantum state, leading to the repulsive interaction. The intermolecular proximity repulsion is typically expressed in the form of 1/r\textsuperscript{n} or exp(b/r). In the field of combustion, we usually use the expression 1/r\textsuperscript{12} \cite{brown-re}. Conversely, when molecules are far apart, the dominance shifts to the interaction of their electric fields. The electric field of a molecule is divided into its intrinsic electric field and its polarized electric field induced by neighboring molecules. Based on the type of interacting electric field, we classify long-range interactions into the electrostatic, dispersion, and induction interactions, creating a second-order tensor interaction space as is illustrated in Fig. \ref{flow}.

\begin{equation}
        \varphi=\varphi_{repul}+\varphi_{elst}+\varphi_{indu}+\varphi_{disp}
        \end{equation}

The electrostatic interaction (Eq. 10) emerges from the interplay of intrinsic electric fields among molecules. The intrinsic electric field of a molecule is typically expanded into various terms, encompassing the intrinsic charge, the intrinsic dipole moment, the intrinsic quadrupole moment, and so forth. Consequently, the electrostatic interaction is often expressed as the interaction between multipole unfolding terms. Given that the contribution of higher-order multipole moments diminishes rapidly at greater intermolecular distances \cite{text}, our focus narrows to the interactions between specific terms, such as the charge term and the dipole moment term.

\begin{equation}
\begin{aligned}
        \varphi_{elst} = & \varphi_{e-e}+\varphi_{e-\mu}+\varphi_{\mu-\mu} \\
        = & +\frac{e_i e_j}{r} \\
        & -\left(\frac{e_i \mu_j}{r^2} \cos \theta_j+\frac{e_j \mu_i}{r^2} \cos \theta_i\right) \\
        & -\frac{\mu_i \mu_j}{r^3}\left(2 \cos \theta_i \cos \theta_j-\sin \theta_i \sin \theta_j \cos \Phi\right)
        \end{aligned}
        \end{equation}
where $\varphi_{e-e}$, $\varphi_{e-\mu}$, $\varphi_{\mu-\mu}$ are the charge-charge, the charge-dipole, and the dipole-dipole interactions, respectively.

The arrangement of charges within the molecule changes in response to the presence of an external electric field, when the molecule is surrounded by other molecules. The deformation of the electron cloud leads to a change in the electric field of the molecule itself, resulting in a polarized electric field. This interaction between the polarized electric field of the molecule and the intrinsic electric field of the surrounding molecules is termed the induction interaction between molecules  (Eq. 11).

\begin{equation}
\begin{aligned}
        \varphi_{indu} = & \varphi_{\alpha e-e}+\varphi_{\alpha e-\mu}+\varphi_{\alpha \mu-e}+\varphi_{\alpha \mu-\mu}  \\
        =& -\left(\frac{\alpha_i e_j^2}{2 r^4}+\frac{\alpha_j e_i^2}{2 r^4}\right) \\
        & -\left(\frac{2 \alpha_i e_j \mu_j}{r^5} \cos \theta_j+\frac{2 \alpha_j e_i \mu_i}{r^5} \cos \theta_i\right)  \\
        & -\left(\frac{\alpha_i \mu_j^2}{r^6}\frac{3 \cos ^2 \theta_j+1}{2}+\frac{\alpha_j \mu_i^2}{r^6}\frac{3 \cos ^2 \theta_i+1}{2}\right)
        \end{aligned}
        \end{equation}
where $\varphi_{\alpha e-e}$, $\varphi_{\alpha e-\mu}$, $\varphi_{\alpha \mu-e}$, $\varphi_{\alpha \mu-\mu}$ are the induced charge-charge, the induced charge-dipole, the induced dipole-charge, and the induced dipole-dipole interactions, respectively.  

Similarly, the dispersion interaction is caused by the polarized electric field of nearby molecules. In contrast to the induction interaction, this kind of interaction occurs between the polarized electric fields of two molecules, which exists between all molecules. The following expression is expressed according to the London dispersion theorem \cite{text},

\begin{equation}
        \varphi_{disp}=-\frac{1.5 \alpha_i \alpha_j}{r^6}\left(\frac{I_i I_j}{I_i+I_j}\right)
        \end{equation}

It can be seen that the electrostatic interaction and the induction interaction contain spatial terms, and their magnitudes are related to the relative orientation of molecules in space. Therefore, polar gases, radicals, ions, and other kind of polar species need to consider the anisotropy of the potential parameters, which mainly depends on the following two factors \cite{text},
        
(i) The electric field generated by the molecules induces the dipoles in this electric field to be regularly arranged.  
       
(ii) The random motion of the molecules attempts to disrupt the ordered arrangement of dipoles, causing the molecular system to tend to a chaotic distribution.  
       
Thus, at lower temperatures, the dipole moments of particles, such as polar molecules, radicals, and ions, exhibit apparent regular arrangements due to the presence of the external electric field, which causes significant polarities of molecules. As the temperature rises, the molecular motion becomes more pronounced, disrupting the ordered arrangement and diminishing the distinctions between nonpolar and polar gases. Therefore, to accurately predict the transport properties of species across a wide range of temperatures and pressures, we introduce the Boltzmann factor to the potential model, taking into account the anisotropy and temperature dependence of potential parameters. 
       
\begin{equation}
\begin{aligned}
        \bar{\varphi} = & \frac{\iiint \varphi \cdot \exp \left(-\varphi /\left(k_B T\right)\right) \cdot \sin \theta_i \sin \theta_j d \theta_i d \theta_j d \Phi}{\iiint \exp \left(-\varphi /\left(k_B T\right)\right) \cdot \sin \theta_i \sin \theta_j d \theta_i d \theta_j d \Phi} \\
        = & \frac{\iiint \varphi\left[1-\frac{\varphi-\bar{\varphi}}{k_B T}+\cdots\right] \cdot \sin \theta_i \sin \theta_j d \theta_i d \theta_j d \Phi}{\iiint\left[1-\frac{\varphi-\bar{\varphi}}{k_B T}+\cdots\right] \cdot \sin \theta_i \sin \theta_j d \theta_i d \theta_j d \Phi} \\
        = & \frac{1}{64 \pi^6} \iiint \varphi \cdot \sin \theta_i \sin \theta_j d \theta_i d \theta_j d \Phi  \\
        & -\frac{1}{k_B T}\frac{1}{64 \pi^6} \iiint \varphi^2 \cdot \sin \theta_i \sin \theta_j d \theta_i d \theta_j d \Phi \\
        & -\frac{1}{k_B T}\left(\frac{1}{64 \pi^6} \iiint \varphi \cdot \sin \theta_i \sin \theta_j d \theta_i d \theta_j d \Phi\right)^2 \\
        & +\cdots
        \end{aligned}
        \end{equation}

Based on the equations provided by Hirschfelder et al. \cite{text} at page 985  (Eq. 14), we derive the weighted electrostatic interaction and the  induction interaction expressions.

\begin{equation}
\begin{aligned}
        \overline{\cos \theta_i} & =\frac{1}{64 \pi^6} \iiint \cos \theta_i \cdot \sin \theta_i \sin \theta_j d \theta_i d \theta_j d \Phi=0 \\
        \overline{\cos ^2 \theta_i} & =\frac{1}{64 \pi^6} \iiint \cos ^2 \theta_i \cdot \sin \theta_i \sin \theta_j d \theta_i d \theta_j d \Phi=\frac{1}{3} \\
        \overline{\left(2 \cos \theta_i \cos \theta_j-\sin \theta_i \sin \theta_j \cos \Phi\right)} & =\frac{1}{64 \pi^6} \iiint\left(2 \cos \theta_i \cos \theta_j-\sin \theta_i \sin \theta_j \cos \Phi\right) \cdot \sin \theta_i \sin \theta_j d \theta_i d \theta_j d \Phi=0 \\
        \overline{\left(2 \cos \theta_i \cos \theta_j-\sin \theta_i \sin \theta_j \cos \Phi\right)^2} & =\frac{1}{64 \pi^6} \iiint\left(2 \cos \theta_i \cos \theta_j-\sin \theta_i \sin \theta_j \cos \Phi\right)^2 \cdot \sin \theta_i \sin \theta_j d \theta_i d \theta_j d \Phi=\frac{2}{3}
        \end{aligned}
        \end{equation}

Hirschfelder et al. \cite{text} have highlighted that the electrostatic interaction is mainly determined by the second term of Eq. 13 (Hirschfelder et al. \cite{text}, page 28). Similarly, the induction interaction is mainly determined by the first term of Eq. 13 (Hirschfelder et al. \cite{text}, page 987). Then we can obtain,

\begin{equation}
\begin{aligned}       
        \bar\varphi_{\text {elst}} & =\bar\varphi_{e-e}+\bar\varphi_{e-\mu}+\bar\varphi_{\mu-u} \\
        & =\frac{e_i e_j}{r}-\left(\frac{e_i^2 \mu_j^2}{r^4} \frac{\overline{\cos ^2 \theta_j}}{k_B T}+\frac{e_j{}^2 \mu_i{}^2}{r^4} \frac{\overline{\cos ^2 \theta_i}}{k_B T}\right)-\frac{\mu_i{}^2 \mu_j{}^2}{r^6} \overline{\frac{\left(2 \cos \theta_i \cos \theta_j-\sin \theta_l \sin \theta_j \cos \phi\right)^2}{k_B T}} \\
        & =\frac{e_i e_j}{r}-\left(\frac{e_i^2 \mu_j^2}{3 k_B T \cdot r^4}+\frac{e_j{}^2 \mu_i{}^2}{3 k_B T \cdot r^4}\right)-\frac{2 \mu_i{}^2 \mu_j{}^2}{3 k_B T \cdot r^6} \\
        \end{aligned}
        \end{equation}

\begin{equation}
\begin{aligned}
        \bar\varphi_{indu} = & \bar\varphi_{\alpha e-e}+\bar\varphi_{\alpha e-\mu}+\bar\varphi_{\alpha \mu-e}+\bar\varphi_{\alpha \mu-\mu}  \\
        =& -\left(\frac{\alpha_i e_j^2}{2 r^4}+\frac{\alpha_j e_i^2}{2 r^4}\right)-\left(\frac{2 \alpha_i e_j \mu_j}{r^5} \overline{\cos \theta_j}+\frac{2 \alpha_j e_i \mu_i}{r^5} \overline{\cos \theta_i}\right) \\
        & -\left(\frac{\alpha_i \mu_j^2}{r^6}\left(\frac{3 \overline{\cos ^2 \theta_j}+1}{2}\right)+\frac{\alpha_j \mu_i^2}{r^6}\left(\frac{3 \overline{\cos ^2 \theta_i}+1}{2}\right)\right) \\
        =& -\left(\frac{\alpha_i e_j{}^2}{2 r^4}+\frac{\alpha_j e_i{}^2}{2 r^4}\right)-\left(\frac{\alpha_i \mu_j{}^2}{r^6}+\frac{\alpha_j \mu_i{}^2}{r^6}\right) \\
        \end{aligned}
        \end{equation}



\begin{equation}
        \bar\varphi_{disp}=-\frac{1.5 \alpha_i \alpha_j}{r^6}\left(\frac{I_i I_j}{I_i+I_j}\right)
        \end{equation}

The complete interaction expression between molecules can be expressed,

\begin{equation}
\begin{aligned}
        \bar{{\varphi}}(r) = & \bar\varphi_{repul}+\bar\varphi_{elst}+\bar\varphi_{indu}+\bar\varphi_{disp}  \\
        = & 4 \varepsilon\left(\frac{\sigma}{r}\right)^{12} \\
        & -\Bigg[-\frac{e_i e_j}{r}+\frac{e_i{}^2 \mu_j{}^2+e_j{}^2 \mu_i{}^2}{3 k_B T \cdot r^4}+\frac{2 \mu_i{}^2 \mu_j{}^2}{3 k_B T \cdot r^6} \\
        & +\left(\frac{\alpha_i e_j{}^2}{2 r^4}+\frac{\alpha_j e_i{}^2}{2 r^4}\right)+\left(\frac{\alpha_i \mu_j{}^2}{r^6}+\frac{\alpha_j \mu_i{}^2}{r^6}\right) \\
        & +\frac{1.5 \alpha_i \alpha_j}{r^6}\left(\frac{I_i I_j}{I_i+I_j}\right)\Bigg]
        \end{aligned}
        \end{equation}

Most of existing potential models determine the potential parameters of pure gas by introducing bath gas and employing the mixing rule to model, which often causes uncertainty of parameters. Thus, we directly calculate potential parameters of the pure gas.

\begin{equation}
        \alpha_i=\alpha_j=\alpha, e_i=e_j=e, \mu_i=\mu_j=\mu, I_i=I_j=I
        \end{equation}

Then Eq. 18 is simplified as follows,

\begin{equation}
\begin{aligned}
        \bar{\varphi}(r) = & 4 \varepsilon\left(\frac{\sigma}{r}\right)^{12}  \\
        & -\Bigg[-\frac{e^2}{r}+\frac{2 e^2 \mu^2}{3 k_B T \cdot r^4}+\frac{2 \mu^4}{3 k_B T \cdot r^6} \\
        & +\frac{\alpha e^2}{r^4}+\frac{2 \alpha \mu^2}{r^6} \\
        & +\frac{0.75 \alpha^2 I}{r^6}\Bigg]
        \end{aligned}
        \end{equation}

\begin{equation}
\begin{aligned}
        \bar{\varphi}(r) = & 4 \varepsilon\Bigg[\left(\frac{\sigma}{r}\right)^{12} + \frac{e^2}{4 \varepsilon \sigma} \frac{\sigma}{r} \\
        & -\left(\frac{2 e^2 \mu^2}{3 k_B T \cdot 4 \varepsilon \sigma^4} + \frac{\alpha e^2}{4 \varepsilon \sigma^4}\right)\left(\frac{\sigma}{r}\right)^4 \\
        & -\left(\frac{2 \mu^4}{3 k_B T \cdot 4 \varepsilon \sigma^6} + \frac{2 \alpha \mu^2}{4 \varepsilon \sigma^6} + \frac{0.75 \alpha^2 I}{4 \varepsilon \sigma^6}\right)\left(\frac{\sigma}{r}\right)^6\Bigg]
        \end{aligned}
        \end{equation}

Next, we introduce a shape correction based on Eq. 21 according \cite{force},

\begin{equation}
\begin{aligned}
        \bar{\varphi}(r) = & 4 \varepsilon\Bigg[(1+2D)\left(\frac{\sigma}{r}\right)^{12} + \frac{e^2}{4 \varepsilon \sigma} \frac{\sigma}{r} \\
        & -\left(\frac{2 e^2 \mu^2}{3 k_B T \cdot 4 \varepsilon \sigma^4} + \frac{\alpha e^2}{4 \varepsilon \sigma^4}\right)\left(\frac{\sigma}{r}\right)^4 \\
        & -\left(\frac{2 \mu^4}{3 k_B T \cdot 4 \varepsilon \sigma^6} + \frac{2 \alpha \mu^2}{4 \varepsilon \sigma^6} + \frac{0.75 \alpha^2 I}{4 \varepsilon \sigma^6}\right)\left(\frac{\sigma}{r}\right)^6\Bigg]
        \end{aligned}
        \end{equation}

The shape correction factor is,

\begin{equation}
        D=A\left[1-\left(\frac{L_2}{L_1}\right)\right]
        \end{equation}
where L1 and L2 are the length and width of the main chain of the molecule respectively, and A is 0.50 for rodlike and -0.25 for platelike molecules.
        
As are shown in Eqs. 21 and 22, the anisotropy and the temperature dependence are inherently considered in the derivation of potential parameters, such as the potential well, molecular diameter, dipole moment, and polarizability, as the Boltzmann factor has been introduced in the weighted potential energy model.

\subsection{Method} \addvspace{10pt}

The development of an accurate potential model is a challenging task, and its precision relies heavily on the potential energy surfaces of intermolecular interactions, apart from the chosen potential functional form. Nowadays, there are many software programs that can perform high-level intermolecular interaction calculations. For example, widely used quantum chemistry software programs like Gaussian and ORCA typically employ the Basis Set Superposition Error (BSSE) \cite{BSSE} correction to determine the potential energy surface between two interacting molecules. However, this method requires additional computational efforts, leading to more than twice the time of the original single-point energy calculation \cite{compare}. Hence, researchers propose the Symmetry-Adapted Perturbation Theory (SAPT) \cite{SAPT} to provide a method for directly calculating the intermolecular potential energy. By applying a perturbation to a system, the intermolecular potential energy can be determined without calculating the total energy of the monomer and the dimer, thus the BSSE correction is not required. Thus, the SAPT method is widely used for the calculation of the noncovalent interaction between molecules. Several programs can perform the SAPT calculations, among which the PSI4 software \cite{psi4} stands out as a commendable choice. PSI4 is not only open-source, free, and fast but also provides a high order of SAPT2+3 accuracy. Therefore, in this paper, we use the PSI4 software to calculate the intermolecular potential energy. 

In PSI4, the calculation accuracy increases with the number of perturbation levels considered,  

SAPT0 $<$ SAPT2 $<$ SAPT2+ $<$ SAPT2+(3) $<$ SAPT2+3.

At the same time, the choice of the basis set also affects the accuracy. Three combinations are recommended for optimal accuracy in the order \cite{order}, 

sSAPT0/jun-cc-pVDZ $<$ SAPT2+/aug-cc-pVDZ $<$ SAPT2+(3)$\delta$MP2/aug-cc-pVTZ. 

Lu \cite{lu} found in tests that SAPT2+(3)$\delta$MP2/aug-cc-pVTZ combination provides accurate results for intermolecular potential energy, with relative errors of only 3\% compared to the accepted high-precision combination CCSD(T)/jun-cc-pVTZ, and absolute errors of less than 0.1 kcal/mol for most systems, but with only one tenth of the calculation time. Therefore, in this paper, we use this combination SAPT2+(3)$\delta$MP2/aug-cc-pVTZ for calculating the intermolecular potential energy. It should be noted that radicals, long-chain molecules can only use the combination SAPT0/jun-cc-pVDZ and sSAPT0/jun-cc-pVDZ due to the special open-shelled structure and excessive size of the molecular system, respectively.

Using the PSI4 software under the above computational settings, we acquired extensive potential energy surface datasets for numerous species. Considering the anisotropy of potential parameters, it was imperative to obtain the full-dimensional potential energy surfaces across all relative orientations. To ascertain the optimal number of orientations, we conducted a test on the direction dependence to evaluate the full-dimensional potential energy surfaces. As is shown in Fig. S1, the convergence of the energy surface was achieved with 432 orientations, and it almost overlaps with that from 864 orientations. Subsequently, as is depicted in Fig. S2, through training the potential energy surface datasets, we attained the Boltzmann-weighted Full-dimensional potential parameters for a number of species, including polar and nonpolar gases, radicals, long-chain molecules, and ions. Since the Boltzmann factor $\exp\left(-\frac{\phi(r)}{k_B T}\right)$ is a function of temperature, the Boltzmann weighting allowed us to concurrently account for the anisotropy and temperature dependence of the potential parameters. These BWF potential parameters can be directly applied to the transport input files of CHEMKIN, Cantera, and other simulation software, as are shown in the Tables S1-S5.

\section{Results and discussion} \addvspace{10pt}

\subsection{Transport property} \addvspace{10pt}

\subsubsection{Common small molecule} \addvspace{10pt}

Before delving into the study of complex molecules, the BWF potential model is verified through transport property predictions, such as the viscosity, mass diffusivity, and thermal conductivity coefficients of small molecules against the TRANLIB library \cite{TRANLIB}, potential parameters in literatures, and the NIST library \cite{NIST}. It is noted that most transport coefficients in present combustion simulations are accomplished from the TRANLIB library and embedded to CHEMKIN \cite{CHEMKIN} and Cantera \cite{Cantera} numerical simulations \cite{brown-re}.

\begin{figure*}[ht!]
    \centering
    \vspace{-0.4 in}
    \includegraphics[width=410pt]{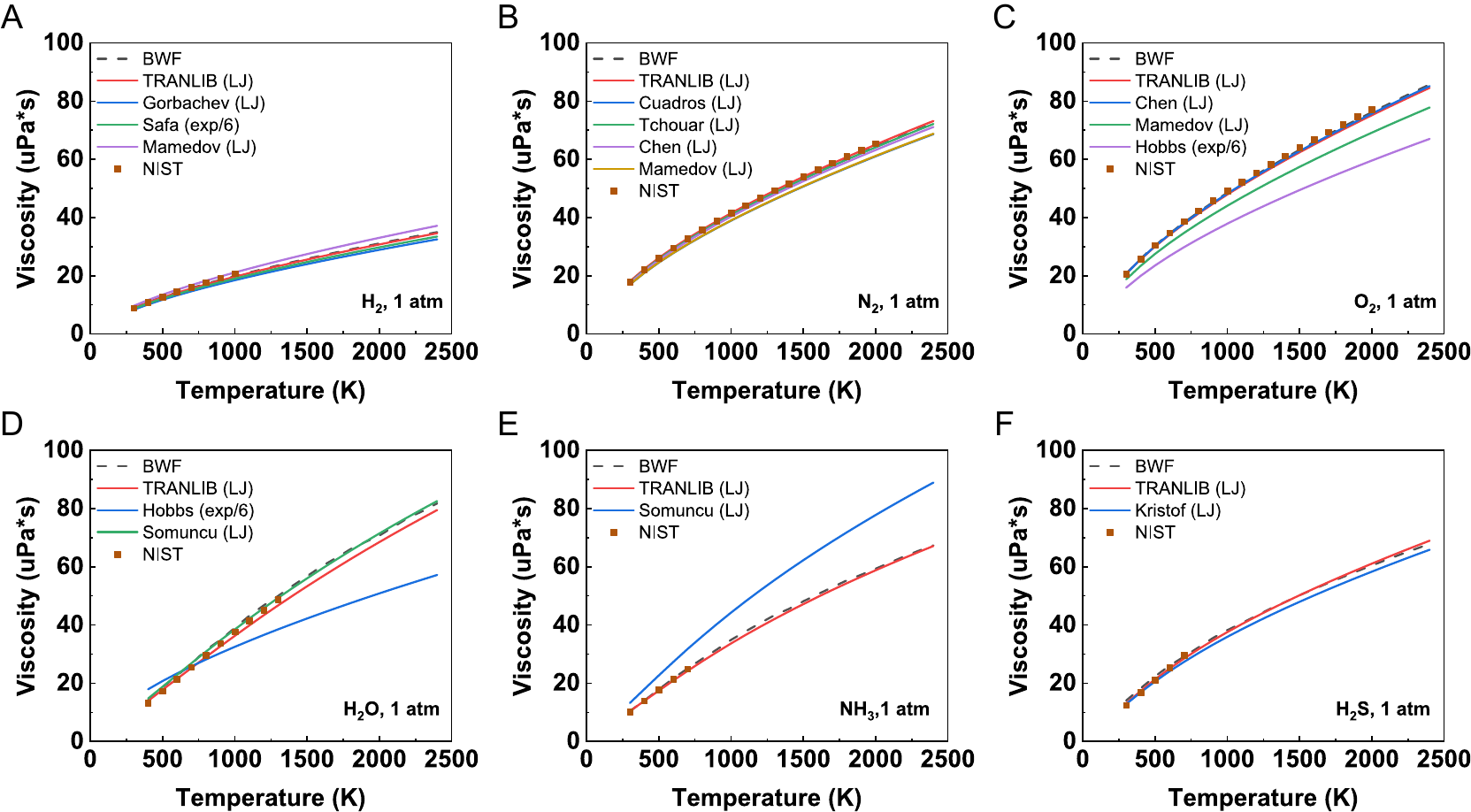}
    \vspace{5 pt}
    \caption{Viscosity coefficients of nonpolar and polar gases at different temperatures (BWF: the Boltzmann-weighted Full-dimensional Potential, LJ: the Lennard–Jones potential, exp/6: the Buckingham potential \cite{buch}) (A) H\textsubscript{2}: Gorbachev \cite{Gorbachev}, Safa \cite{Safa}, Mamedov \cite{Mamedov}, NIST \cite{NIST}; (B) N\textsubscript{2}: Cuadros \cite{Cuadros}, Tchouar \cite{Tchouar}, Chen \cite{Chen}, Mamedov \cite{Mamedov}, NIST \cite{NIST}; (C) O\textsubscript{2}: Chen \cite{Chen}, Mamedov \cite{Mamedov}, Hobbs \cite{Hobbs}, NIST \cite{NIST}; (D) H\textsubscript{2}O: Hobbs \cite{Hobbs}, Somuncu \cite{Somuncu}, NIST \cite{NIST}; (E) NH\textsubscript{3}: Somuncu \cite{Somuncu}, NIST \cite{NIST}; (F)H\textsubscript{2}S: Kristof \cite{Krist}, NIST \cite{NIST}.}
    \label{vis-T}
    \end{figure*}

For nonpolar molecules such as H\textsubscript{2}, N\textsubscript{2}, and O\textsubscript{2}, by incorporating the potential parameters into Eqs. 1-3, we calculate the coefficients of the viscosity, mass diffusivity, and thermal conductivity for each nonpolar gas across a wide temperature range of 300-2400 K. Fig. \ref{vis-T} exhibits the viscosity coefficients with temperature. The BWF and the TRANLIB (LJ) viscosity coefficients possess errors less than 1\% from experimental values for H\textsubscript{2}, N\textsubscript{2}, and O\textsubscript{2}, exhibiting excellent predictabilities. While the Gorbachev (LJ) model creates a relative error of 9.5\% for H\textsubscript{2}, the Cuadros (LJ) and Mamedov (LJ) models maintain errors of 6\% for N\textsubscript{2}. Most LJ or exp/6 potential models in literatures fail to accurately describe intermolecular interactions of O\textsubscript{2} due to its unique open-shell energy structure. For instance, the Hobbs (exp/6) model displays a relative error of 23\%. Similarly, as are shown in Figs. S4 and S5, the BWF potential model shows the best consistency with the TRANLIB library and experimental values for mass diffusivity and thermal conductivity coefficients.

As to polar molecules such as H\textsubscript{2}O, NH\textsubscript{3}, and H\textsubscript{2}S, the BWF and the TRANLIB (LJ) viscosity coefficients exhibit errors within 5\% from experimental values in Fig. \ref{vis-T}. Unfortunately, other LJ or exp/6 potential models in literatures remain larger uncertainties. For example, the Somuncu (LJ) model shows relative errors of 10\% and 28.3\% for H\textsubscript{2}O and NH\textsubscript{3}, respectively. Moreover, for the mass diffusivity and thermal conductivity coefficients of polar gases, the BWF potential model also reveals the highest accuracy with experimental values, as are shown in Figs. S4 and S5.

\begin{figure*}[ht!]
        \centering
        \vspace{-0.0 in}
        \includegraphics[width=380pt]{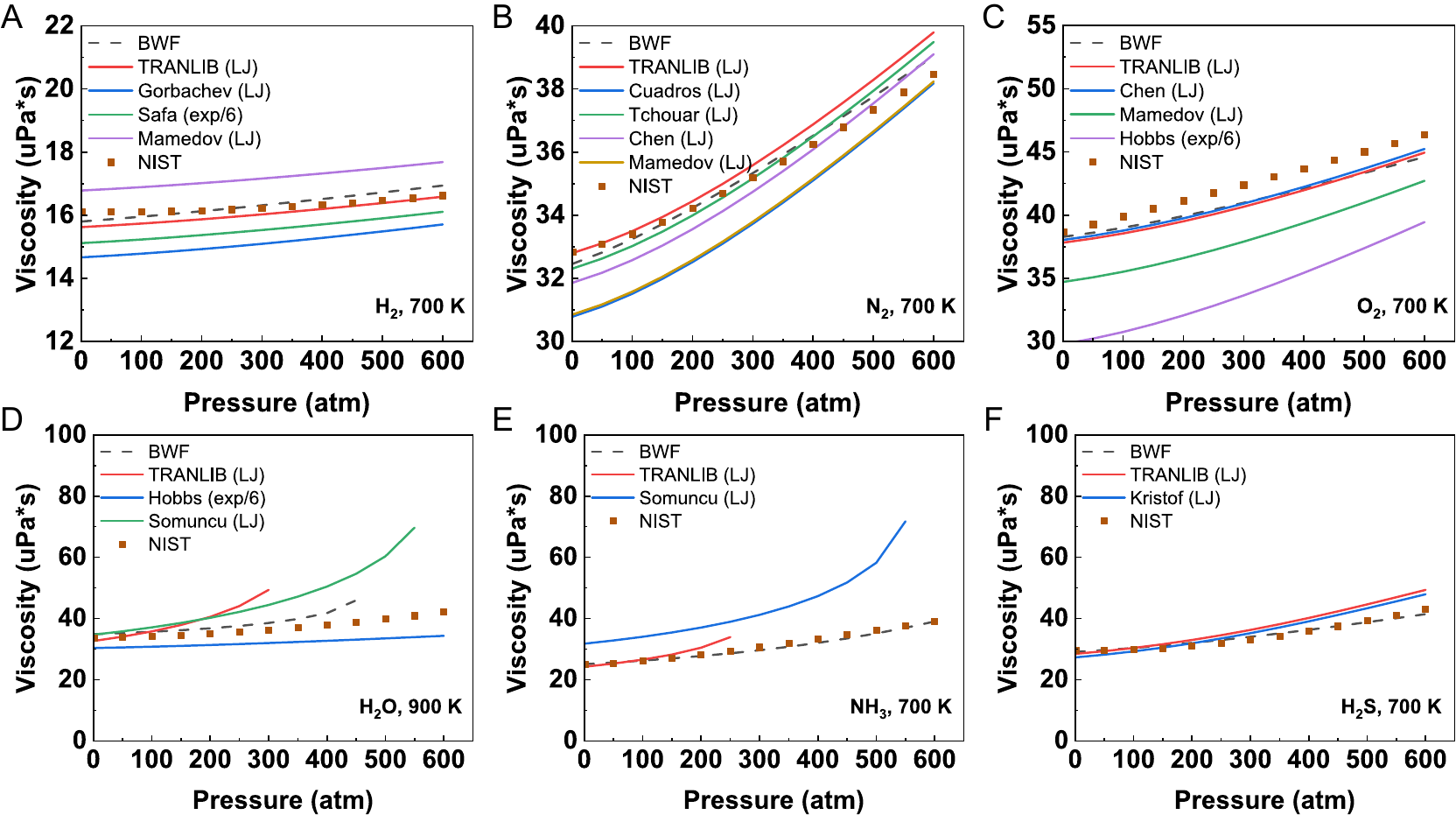}
        \vspace{5 pt}
        \caption{Viscosity coefficients of nonpolar and polar gases at different pressures (BWF: the Boltzmann-weighted Full-dimensional Potential, LJ: the Lennard–Jones potential, exp/6: the Buckingham potential \cite{buch}) (A) H\textsubscript{2}: Gorbachev \cite{Gorbachev}, Safa \cite{Safa}, Mamedov \cite{Mamedov}, NIST \cite{NIST}; (B) N\textsubscript{2}: Cuadros \cite{Cuadros}, Tchouar \cite{Tchouar}, Chen \cite{Chen}, Mamedov \cite{Mamedov}, NIST \cite{NIST}; (C) O\textsubscript{2}: Chen \cite{Chen}, Mamedov \cite{Mamedov}, Hobbs \cite{Hobbs}, NIST \cite{NIST}; (D) H\textsubscript{2}O: Hobbs \cite{Hobbs}, Somuncu \cite{Somuncu}, NIST \cite{NIST}; (E) NH\textsubscript{3}: Somuncu \cite{Somuncu}, NIST \cite{NIST}; (F)H\textsubscript{2}S: Kristof \cite{Krist}, NIST \cite{NIST}.}
        \label{vis-P}
        \end{figure*}

Unfortunately, in most cases, the pressure dependence is not considered in the transport property computations, while it becomes non-negligible for real fluids such as supercritical fluid at high pressures. Therefore, for real-fluid computations, we corrected the pressure dependences of viscosity (Fig. \ref{vis-P}), mass diffusivity (Fig. S6), and thermal conductivity (Fig. S7) coefficients based on the Enskog theory \cite{text} (Eqs. 5-8). It shows that the corrected transport coefficients exhibit significant pressure dependences across a wide pressure range of 1-600 atm. For nonpolar gases, such as H\textsubscript{2}, N\textsubscript{2}, and O\textsubscript{2}, the relative errors of the BWF and the TRANLIB (LJ) viscosity coefficients are both around 1-3\%. However, the relative errors of the Mamedov (LJ) model reach 6\% for H\textsubscript{2} and N\textsubscript{2}, and the Hobbs (exp/6) viscosity coefficients for O\textsubscript{2} reach an error of 22\%, which beyond the acceptable prediction error of 5\% \cite{brown-re}.
 
As is shown in Figs. \ref{vis-P} D, E, and F, for polar molecules, such as H\textsubscript{2}O, NH\textsubscript{3}, and H\textsubscript{2}S, even though these LJ potential models take into account the electrostatic interaction between dipole moments (the Stockmayer potential), their predictive ability remains inadequate. For example, for NH\textsubscript{3}, the prediction error of the BWF viscosity coefficients is around 3\%, while the error of the TRANLIB (LJ) model reaches 15.6\% at 250 atm, and other LJ potentials exhibit even higher errors. Furthermore, the several LJ potentials in Figs. \ref{vis-P} D and E significantly diverge beyond 250 atm. Therefore, the BWF model exhibit a higher prediction accuracy for nonpolar and polar molecules than various LJ models in literatures as the former takes various intermolecular interactions in account comprehensively.
                            
\subsubsection{Radicals, long-chain molecules, and ions} \addvspace{10pt}

It is noted that the transport coefficient data in the TRANLIB library is mainly fitted from experimental measurements \cite{brown-re}. Therefore, it agrees well with the experimental transport coefficients of nonpolar and polar molecules. The successful agreement between our theoretical BWF potential model with the TRANLIB library and experimental values in the broad range of temperatures and pressures (300-2400 K and 1-600 atm), confirms its applicability for transport property predictions. The BWF potential parameters of nonpolar and polar gases are attached in the Tables S1 and S2. Unfortunately, for radicals, long-chain molecules, and ions, where the experimental data is rarely accessed, the TRANLIB library will lose its fitting accuracy.

\begin{figure*}[ht!]
        \centering
        \vspace{-0.0 in}
        \includegraphics[width=300pt]{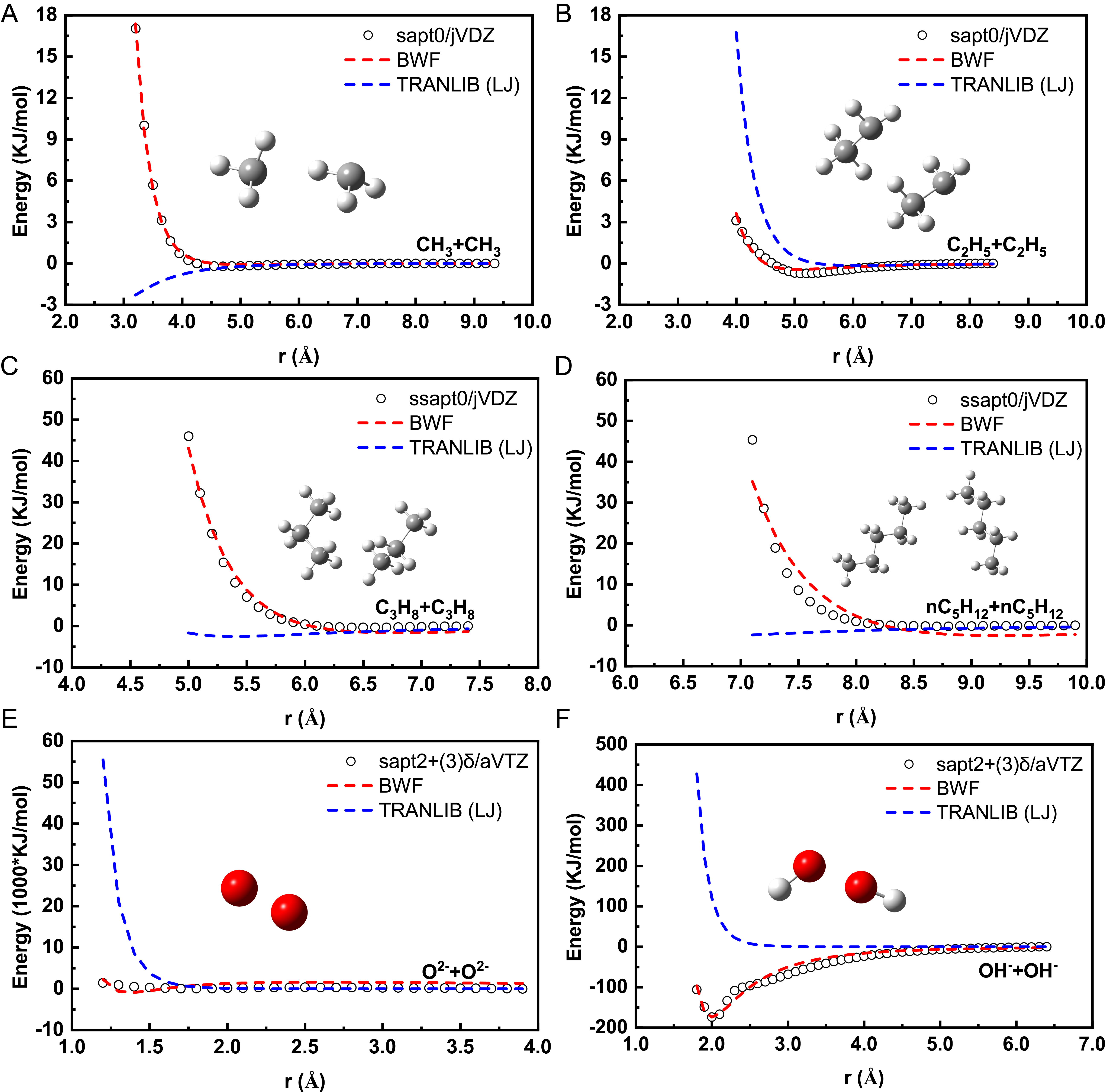}
        \vspace{5 pt}
        \caption{The Boltzmann-weighted Full-dimensional potential energy surfaces for radicals, long-chain molecules, and ions (o: the SAPT computations, sapt0/jVDZ: SAPT0/jun-cc-pVDZ combination in PSI4 software, ssapt0/jVDZ: sSAPT0/jun-cc-pVDZ combination in PSI4 software, sapt2+(3)$\delta$/aVTZ: SAPT2+(3)$\delta$MP2/aug-cc-pVTZ combination in PSI4 software) (A) CH\textsubscript{3}; (B) C\textsubscript{2}H\textsubscript{5}; (C) C\textsubscript{3}H\textsubscript{8}; (D) nC\textsubscript{5}H\textsubscript{12}; (E) O\textsuperscript{2-}; (F) OH\textsuperscript{-}.}
        \label{curve}
        \end{figure*}

Radicals own the high reactivity, short life time, and large polarity, resulting in difficulties in experimental measurements and rigorous calculations. As are shown in Figs. \ref{curve} A and B, the energy potential surfaces of CH\textsubscript{3} and C\textsubscript{2}H\textsubscript{5} are plotted by using the TRANLIB (LJ) and the BWF potential parameters, and compared with the theoretical trajectory from high-level quantum chemistry computations. It is evident that the LJ potential model cannot accurately describe the potential energy surfaces especially the repulsive walls \cite{Jasper1}. In contrast, the BWF potential model, after considering comprehensive intermolecular interactions and introducing more potential parameters, effectively improves the repulsive walls of the potential surfaces, closely aligning with the theoretical energy-potential trajectory. Since there are a large number of radicals in the combustion system, it is particularly important to improve their transport properties. We have calculated the potential parameters of a large number of radicals, CH, CH\textsubscript{2}, CH\textsubscript{3}, C\textsubscript{2}H\textsubscript{3}, C\textsubscript{2}H\textsubscript{5}, CHO, H, and other radicals, as are shown in the Table S3.

As the main components of fuels, the transport properties of long-chain molecules are pivotal for combustion simulations. Due to their long-chain structure, treating these molecules as simple point particles introduces a large amount of errors into transport properties \cite{Jasper1, Violi1, Violi2}. Therefore, we introduce a shape correction based on Eqs. 21 and 22 in the section of Materials and Methods. As are shown in Figs. \ref{curve} C and D, the potential energy surfaces of the TRANLIB (LJ) and the BWF potential model show significant differences. The LJ potential surface without shape correction deviates significantly from the Symmetry-Adapted Perturbation Theory calculations, and seriously underestimates the diameter of the molecules, only estimating the propane molecular diameter of 4.81 $\AA$ and the pentane molecular diameter of 5.596 $\AA$. Therefore, taking into account the internal configuration of long-chain molecules and shape correction is crucial for the accurate description of the interactions between long-chain molecules. Using the corrected potential model, we trained the dataset of long-chain molecules and derived the corresponding potential parameters, as are shown in the Table S4.
        
Traditional potential models, such as the LJ potential model, often overlook the Coulombic electrostatic and the ion-induced interactions and only handle the simplest uncharged molecules. However, with the continuous development and progress of the combustion field, such as the plasma-assisted combustion, how to deal with the charged particles in the combustion system has become an important topic to be concerned about. The BWF potential model, in addition to the basic repulsive and dispersion interactions, also takes into account the electrostatic and induction interactions. It can describe the interaction of a variety of particles, including uncharged particles as well as charged particles. As are shown in Figs. \ref{curve} E and F, the BWF potential model can accurately reproduce the interaction between charged particles, making up for the shortcoming that the traditional LJ potential model cannot deal with charged particles. The potential parameters of OH\textsuperscript{-} and O\textsuperscript{2-}are shown in the Table S5.
                        
\subsection{Combustion characteristics} \addvspace{10pt} 

The model validation above has successfully confirmed that in a wide range of temperatures and pressures (300-2400 K and 1-600 atm), the BWF potential model exhibits comprehensive and accurate predictions of transport properties of nonpolar and polar molecules, long-chain molecules, radicals, and ions simultaneously, while other models in literatures do not possess the similar versatility and accuracy. Therefore, the BWF potential model is applied in the real-fluid simulation away from the ideal-gas equilibrium, such as supercritical fluid, high-enthalpy fluid, plasma interactions.

The supercritical combustion, known for its high thermodynamic efficiency and low emissions, has gained significant attention, particularly in advanced engines and gas turbines. Notably, the Raptor engine of the SpaceX Starship was operated in the supercritical combustion region using a CH\textsubscript{4}/O\textsubscript{2} mixture at pressures exceeding 300 atm \cite{spacex}. However, a number of problems, such as engine misfires and malfunctions (3 Raptor engines misfired during the first launch of the Starship), flame instability was found in the Raptor engine operation, implying that our current research on the supercritical combustion is still insufficient, and it is necessary to carefully consider the real-fluid behavior in extreme combustion conditions like the supercritical combustion. Therefore, this work investigates combustion characteristics such as the laminar flame speed and the flame extinction limit of methane at atmospheric and high pressures. In addition, in order to explore the influence of real-fluid effects on fuels with different carbon-chain lengths, we have also conducted insightful studies on dimethyl ether (DME) and n-heptane.

The laminar flame speeds and the flame extinction limits of fuels are numerically calculated using the PREMIX and the OPPDIF modules of the CHEMKIN package, respectively. Various detailed and reduced chemical kinetic models are tested to validate the experimental flame laminar speeds and extinction limits. The GRI3.0 mechanism \cite{GRI30}, the HPmech-1.3 mechanism \cite{HPmech}, and the simplified n-heptane mechanism \cite{CKW} by the PFA method \cite{PFA} are employed to simulate and predict methane, dimethyl ether, and n-heptane fuels, respectively.

\subsubsection{The flame extinction limit}

\begin{figure}[ht!]
    \centering
    \includegraphics[width=200pt]{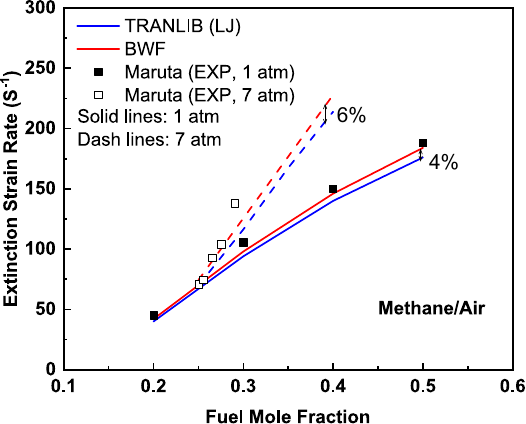}
    \vspace{5 pt}
    \caption{The flame extinction limits of the methane /air at an initial fuel temperature of 550 K, an initial air temperature of 300 K, and initial pressures of 1 and 7 atms with comparisons to the experimental data: Maruta \cite{EL-CH4}.}
    \label{met-EL}
    \end{figure}

\begin{figure}[ht!]
    \centering
    \includegraphics[width=210pt]{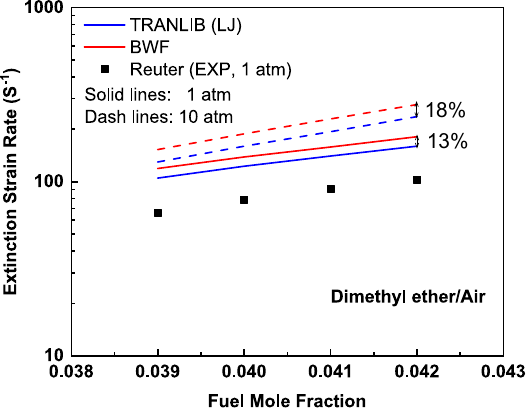}
    \vspace{5 pt}
    \caption{The flame extinction limits of the dimethyl ether /air at an initial fuel temperature of 550 K, an initial air temperature of 300 K, and initial pressures of 1 and 10 atms with comparisons to the experimental data: Reuter \cite{EL-DME}.}
    \label{DME-EL}
    \end{figure}

\begin{figure}[ht!]
    \centering
    \includegraphics[width=210pt]{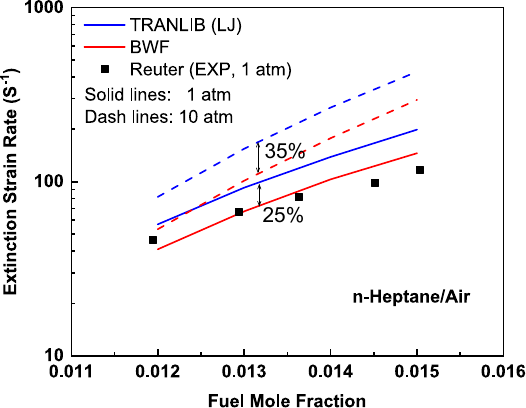}
    \vspace{5 pt}
    \caption{The flame extinction limits of the n-heptane/air at an initial fuel temperature of 550 K, an initial air temperature of 300 K, and initial pressures of 1 and 10 atms with comparisons to the experimental data: Reuter \cite{EL-C7H16}.}
    \label{C7H16-EL}
    \end{figure}

Three Raptor engines of the Starship misfired during the first launch of the interplanetary spacecraft, and our understanding of the flame extinction characteristics is still incomplete. We have investigated the flame extinction limits including methane, DME, and n-heptane, as are shown in Figs. \ref{met-EL}, \ref{DME-EL}, and \ref{C7H16-EL}, respectively. The axial velocity gradient at the oxidizer side is defined as a global strain rate \cite{ziyu}, $a_{ext} = 2U_o/L[1 + ( U_f/U_o )( \rho_f/\rho_o )^{1/2}]$, where $U_o$ and $U_f$ are the flow velocities of oxidizer and fuel streams, and $\rho_o$ and $\rho_f$ the densities of oxidizer and fuel streams, respectively. Compared with the experimental data, the predictions of the BWF and the TRANLIB (LJ) extinction limits for methane are close to each other, showing relative prediction deviations of 5-15\% with experimental measurements. However, for the long-chain molecules such as n-heptane, the BWF potential model exhibits the prediction error level of 10\% from the experimental data, while the TRANLIB (LJ) model possesses a significant uncertainty above 50\% from experiments. As the flame extinction is highly relevant to transport properties of molecules through the extinction strain rate \cite{Wang3,Wang4}, the accurate computations of transport properties suing the BWF potential model show a significant improvement in the flame extinction predictions than LJ models, especially for long-chain molecules where shape corrections are not considered in LJ models in literatures. This underscores the importance of considering the molecular configuration and anisotropy to enhance the prediction accuracy of transport and combustion properties. In addition, the BWF potential model may makes larger difference at higher pressures for flame extinction predictions, while unfortunately, higher pressure data of flame extinction limits are not available in literatures for the model validation.

\clearpage

\subsubsection{The laminar flame speed}

\begin{figure}[ht!]
    \centering
    \includegraphics[width=192pt]{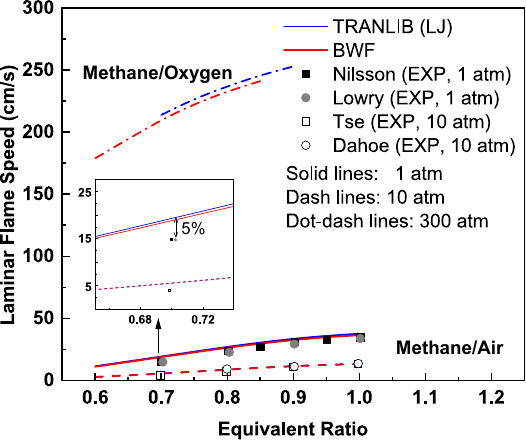}
    \vspace{5 pt}
    \caption{The laminar flame speeds of methane at an initial temperature of 300 K and pressures of 1 atm (methane/air), 10 atm (methane/air), and 300 atm (methane/oxygen) with comparisons to the experimental data: Nilsson \cite{Nilsson}, Lowry \cite{Lowry}, Tse \cite{Tse}, Dahoe \cite{dahoe}.}
    \label{met-SL}
    \end{figure}

\begin{figure}[ht!]
    \centering
    \includegraphics[width=192pt]{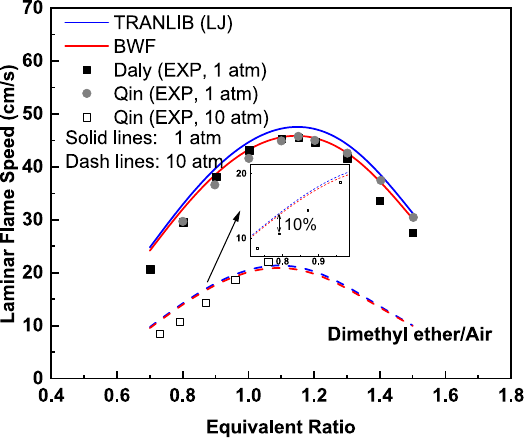}
    \vspace{5 pt}
    \caption{The laminar flame speeds of the dimethyl ether/air at an initial temperature of 300 K and pressures of 1 and 10 atms with comparisons to the experimental data: Daly \cite{Daly}, Qin (1 atm) \cite{Qin}, Qin (10 atm) \cite{Qin}.}
    \label{DME-SL}
    \end{figure}

\begin{figure}[htbp]
    \centering
    \includegraphics[width=192pt]{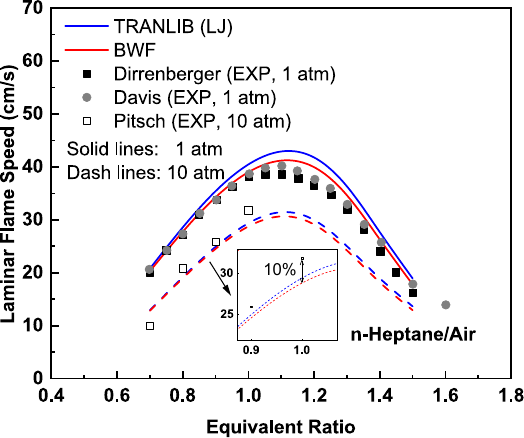}
    \vspace{5 pt}
    \caption{The laminar flame speeds of the n-heptane/air at 300 K and 1 atm,373 K and 10 atm with comparisons to the experimental data: Dirrenberger \cite{Dirrenberger}, Davis \cite{Davis}, Pitsch \cite{Pitsch}.}
    \label{C7H16-SL}
    \end{figure}

Fig. \ref{met-SL} depicts the laminar flame speed versus the equivalent ratio for the methane/air mixture at 1 and 10 atms and the methane/oxygen mixture at 300 atm. It shows that the predicted laminar flame speeds by using both the BWF and the TRANLIB (LJ) method closely match the experimental data, with relative errors around 5\% at 1 and 10 atms. Notably, the BWF potential model aligns slightly better with experimental data at 10 atms than the TRANLIB (LJ) method. While at 300 atm, the prediction discrepancy between these two models was amplified. Moreover, the laminar flame speeds of DME and n-heptane are also compared in Figs. \ref{DME-SL} and \ref{C7H16-SL}, the BWF laminar flame speed is in better agreement with the experimental data than the TRANLIB (LJ) model by up to 10\% at both 1 and 10 atms. Therefore, it indicates a high accuracy of the BWF method for predicting laminar flame speeds, especially for long-chain fuels. 

It is seen that the real-fluid potential model plays a larger impact on correcting the flame extinction limit than the laminar flame speed due to a higher transport relevance. Unfortunately, there is a lacking of experimental data for the measurement of laminar flame speeds and extinction limits above 50 atm to validate the BWF model at high pressures. 

\clearpage


\section{Conclusion} \addvspace{10pt}

Numerical modeling stands as a pivotal instrument for in-depth study of extreme combustion conditions such as the supercritical combustion. Our research focuses on elucidating transport properties, particularly the mass diffusivity, viscosity, and thermal conductivity, which are key factors that have an influence on combustion simulations. The intermolecular interactions need to be considered in the transport modeling, especially in cases where the real-fluid effects are exacerbated. To meet this requirement, we construct a second-order tensor interaction space, the Boltzmann-weighted Full-dimensional potential model, which is adept at accommodating a wide variety of species, including common nonpolar and polar gases, radicals, long-chain alkanes, and ionic species. The model cleverly captures diverse intermolecular interactions, providing key potential parameters, such as the potential well, molecular diameter, dipole moment, and polarizability. No extra errors are introduced for potential parameters by the mixing rules without introducing the bath gases to the systems. Moreover, the BWF potential model contains more parameters and describes the potential energy surface more accurately, effectively improving the problem that LJ model over-steeply predicts the repulsive walls. On this basis, its versatility extends to incorporating anisotropy and temperature dependence in potential parameters through Boltzmann weighting across various spatial orientations.

We conduct the high-level Symmetry-Adapted Perturbation Theory calculations using the PSI4 software, yielding a rich array of potential energy surface datasets. Through careful training on the dataset exceeding $5*10^6$ data, we successfully delineate the Boltzmann-weighted Full-dimensional potential parameters for the common nonpolar and polar gases, radicals, long-chain alkanes, and ions. Then, we use the BWF potential model to provide the transport coefficients over a wide range of temperatures and pressures (300-2400 K and 1-600 atm). These BWF transport properties are compared with the LJ transport properties in literatures and the experimental measurements. Encouragingly, the relative errors of the BWF transport properties hover around 1\% for nonpolar gases and about 5\% for polar molecules, which beyond the acceptable prediction error of 5\%. Expanding the application of the BWF potential model to radicals, long-chain molecules, and ions, the model accurately reproduces the complex interactions between various particles. It effectively improves the problems of the LJ model, including the improper description of repulsive walls, the under-prediction of the diameters of long-chain molecules, and the unavailability of charged particles. Subsequently, we integrate the model into the combustion simulation to calculate the fundamental combustion characteristics, such as the laminar flame speeds and flame extinction limits for methane, dimethyl ether, and n-heptane at high and low pressures. The results confirm that the BWF model is able to accurately predict the combustion characteristics under various complex operating conditions. Notably, for long-chain alkanes, such as n-heptane, the BWF model can accurately improve the prediction of flame extinction limits, and differs by up to 35\% from the LJ model between 1-10 atm.

In conclusion, this work marks a significant advance in the real-fluid modeling. The development of a Boltzmann-weighted Full-dimensional potential model has succeeded in obtaining and validating a variety of potential parameters, which can be used directly as transport input files for the software of the combustion simulation, such as CHEMKIN and Cantera. Besides, a series of transport properties and combustion characteristics of fuels are obtained, which provides strong supports for the development in the field of combustion.

\clearpage

\acknowledgement{Declaration of competing interest} \addvspace{10pt}

The authors declare that they have no known competing financial interests or personal relationships that could have appeared to influence the work reported in this paper.

\acknowledgement{Acknowledgments} \addvspace{10pt}

The work was funded by the National Key Research and Development Program (2023YFE0120900).

\acknowledgement{Supplementary material} \addvspace{10pt}

Supplementary material is submitted along with the manuscript.

 \footnotesize
 \baselineskip 9pt

\bibliographystyle{pci}
\bibliography{PCI_LaTeX}

\newpage

\small
\baselineskip 10pt

\end{document}